\documentclass[12pt, a4paper]{article}
\usepackage{times,epsfig,makeidx,amsfonts,amsmath}
\usepackage{graphicx,amsbsy,amssymb,multicol,rotating,multirow}
\usepackage[round]{natbib}
\usepackage{graphicx,ifthen,calc,tabularx}
\usepackage{xspace}
\usepackage{float}
\usepackage{makecell,booktabs}
\usepackage{url}
\def\bm{\boldsymbol}
\def\N{\mbox{N}}

\renewcommand{\baselinestretch}{1.5}

\newtheorem{algorithm}{Algorithm}

\begin{document}

\title{Adaptive MC$^3$ and Gibbs Algorithms for Bayesian Model Averaging in Linear Regression Models}
\author{ Demetris  Lamnisos, Jim E. Griffin and Mark F.J.~Steel
\thanks{Cyprus University of Technology, University of Kent and University of Warwick }
}

\maketitle

\begin{abstract}
The MC$^3$ \citep{Madigan.York:95} and Gibbs \citep{George.McCullogh:97} samplers are the most widely implemented algorithms for Bayesian Model Averaging (BMA) in linear regression models. These samplers draw a variable at random in each iteration
using uniform selection probabilities and then propose to update that variable. This may be computationally inefficient if the number of variables is large and many variables are redundant. In this work, we introduce adaptive versions of these samplers that retain their simplicity in implementation  and reduce the selection probabilities of the many redundant variables. The improvements in efficiency for the adaptive samplers are illustrated in real and simulated datasets. \\
\\
{\it Keywords}: Adaptive MCMC; Gibbs sampler; Model uncertainty; Variable selection
\end{abstract}

\section{Introduction}

The growing availability of datasets with large number of regressors has lead to an increased interest in incorporating model uncertainty in inference and decision problems. We consider the problem of model uncertainty in a linear regression model with $n$ observations of a response variable in $\bm{y}=(y_1,\ldots,y_n)'$ and a large number of $p$ potential predictors.
The vector of indicator variables $\bm{\gamma}=(\gamma_1,\ldots,\gamma_p)$ is introduced to represent inclusion $(\gamma_i=1)$ or exclusion $(\gamma_i=0)$ of the $i-$th regression leading to model size $p_{\gamma}=\sum_{i=1}^p\gamma_i$. The normal linear regression model conditional on $\bm{\gamma}$ is expressed as
\[\bm{y}|a,\bm{\beta}_{\gamma},\tau,\bm{\gamma}\sim N(\alpha\mathbf{1}+\bm{X}_\gamma{\bm{\beta}}_\gamma, \tau \mathbf{I}_n ),\]
where the design matrix $\bm{X}_\gamma$ contains the measurements of the $p_\gamma$ included variables, $\mathbf{1}$ represents an $(n\times 1)-$dimensional vector of ones, $\mathbf{I}_n$ is the identity matrix of order $n$, $\alpha$ is the intercept, $\bm{\beta}_{\gamma}$ represents the regression coefficients and $\tau$ is the error variance.

Under the most commonly used prior structures, the marginal likelihood has an analytical expression in normal linear regression models, facilitating the computation of the posterior distribution over models. Bayesian Model Averaging (BMA) utilizes this posterior distribution to incorporate model uncertainty in posterior inferences \citep{Hoeting.Madigan:99}. For a specific quantity of interest $\Delta$, the posterior distribution of $\Delta$ under BMA is a mixture of the posterior distribution of $\Delta$ under each model weighted by the posterior model distribution.

When the number of variables $p$ is greater than 25-30, enumeration of all possible models is generally intractable and sampling methods like  Markov chain Monte Carlo
(MCMC)  are often used to explore the model space. These methods provide  a dependent sample of models from the posterior model distribution which is used to
approximate BMA through ergodic averages. The two most widely implemented MCMC samplers for those problems are the MC$^3$ \citep{Madigan.York:95} and Gibbs \citep{George.McCullogh:97} samplers. These samplers draw a variable at random in each iteration using uniform selection probabilities and then propose either to add or delete that variable from the current model of the chain. However, in the large $p$ setting there are often many redundant variables  and the uniform variable selection probabilities then cause a reduction in the efficiency of the algorithms because computational time is wasted in rejecting poor proposals. The design of a new proposal that automatically identifies the redundant variables during the run of the sampler and assigns  much lower selection probabilities to those variables could considerably improve the efficiency of those algorithms in the large $p$ setting.

There has been an interest recently in adaptive MCMC methods that attempt to improve the proposal distribution in a Metropolis-Hastings step during the run of the sampler using information contained in the current sample \citep{haario01,Atchade.Rosenthal:05,Roberts.Rosenthal:07,Roberts.Rosenthal:09,andrieuthoms08}. \cite{LatRos10} develop a class of adaptive samplers that adapt the coordinate selection probabilities of a Gibbs sampler and also study the ergodicity of those samplers. These adaptive MCMC samplers are quite promising in improving efficiency, although standard MCMC theory for the ergodicity of those samplers does not apply. The potential gains from the implementation of adaptive MCMC methods in BMA  for linear regression models have been recognised and there has been a fast growing literature proposing adaptive  MCMC algorithms 
\citep{nottkohn05,Peltola.Martttinen:12,Lamnisos.Griffin:12}.

In this work, we develop adaptive versions of MC$^3$ and Gibbs samplers that adapt the variable selection probabilities in such a way that redundant variables are assigned
lower selection probabilities. As the samplers run, we progressively learn through some basic descriptive sample measures which variables tend to be
redundant and we utilize this information to adapt the variable selection probabilities. More specifically, we propose to periodically update the variable selection probabilities by computing a weighted version of those descriptive sample  measures. These new adaptive samplers automatically decrease the selection probabilities of the many
redundant variables. Thus, we can avoid the computational burden of proposing many poor proposals and  explore the posterior model distribution more efficiently. Moreover, these adaptive algorithms are easy to implement because a single step is added in the simple MC$^3$ and Gibbs samplers. Finally, ergodicity results are proved for those adaptive samplers and we provide a recommended sampler for the  applied user.

The article is organised as follows: Section 2 describes the MCMC algorithms for BMA in linear regression model while Section 3 introduces adaptive MCMC algorithms for BMA in a linear regression and also examines the ergodicity of those adaptive algorithms. The adaptive algorithms are applied to simulated and real datasets in Section 4 and conclusions and recommendations are given in Section 5. Code in Matlab (along with some instructions and the real data sets used in the paper) is freely available at {\url {http://www.warwick.ac.uk/go/msteel/steel_homepage/software/supp_admc3_gibbs.zip}}.


\section{MCMC Algorithms for BMA in the Linear Regression Model}
The Bayesian approach to model uncertainty proceeds by placing a prior distribution on the intercept $\alpha$, the regression coefficients $\bm{\beta}_{\gamma}$,
the error variance $\tau$ and the model $\bm{\gamma}$. A quite common choice of priors in BMA for linear regression is the following
\begin{align*}
\pi(\alpha)   & \propto  1 \\
\pi(\tau)     & \propto  \frac{1}{\tau} \\
\bm{\beta}_{\gamma}|\bm{\gamma},\tau,g & \sim  \N(\bm{0},\tau g(\bm{X}_\gamma'\bm{X}_\gamma)^{-1})\\
\pi(\bm{\gamma}|w) & =  w^{p_\gamma}\;(1-w)^{p-p_\gamma}\\
w             &\sim  \mbox{Beta}(b,c),
\end{align*}
where  the hyperparameters $b$ and $c$ are chosen such that the prior mean of model size $\mathbb{E}(p_\gamma)=\kappa$ and $\mbox{Var}(p_\gamma)=2\kappa(p-\kappa)/p.$
The Benchmark $g-$prior ($g-$BRIC) and the Hyper$-g/n$ prior are the two choices used here for the single parameter $g$. The $g-$BRIC prior, introduced
by \cite{Fernandez.Steel:01}, sets $g=\max\{n,p^2\}$ while the Hyper$-g/n$ prior assigns the following hyperprior to $g$
\[\pi(g)=\frac{a-2}{2n}\left(1+\frac{g}{n}\right)^{-a/2}, \, \mbox{with}\,\, a>2,\]
as proposed by \cite{Liang.Paulo:08}.
The hyperprior on $g$ allows for the data to influence the inference about $g$ and makes the analysis more robust with respect to the assumptions on $g$. The Hyper$-g/n$ prior (with $a=3$) is one of the two priors on $g$ recommended by \cite{Ley.Steel:12} who extensively examine the performance of various priors on $g$ in the context of simulated and real data.

This choice of  priors for the model specific parameters results in an analytical expression
for the marginal likelihood $\pi(\bm{y}|\bm{\gamma},g)$ of model
$\bm{\gamma}$ given by
\begin{equation*}
\pi(\bm{y}|\bm{\gamma},g)\propto
\left(\frac{1}{1+g}\right)^{p_\gamma/2}\,\left(\tilde{\bm{y}}'\tilde{\bm{y}}-\frac{g}{1+g}\bm{y}'
\bm{X}_\gamma(\bm{X}_\gamma'\bm{X}_\gamma)^{-1}\bm{X}_\gamma'\bm{y}\right)^{-(n-1)/2},\label{eqn:marginal_likelihood_improper}
\end{equation*}
where $\tilde{\bm{y}}=\bm{y}-\bar{y}\mathbf{1}$ and $\bar{y}$ is the
mean of the response $\bm{y}$. This analytical expression for the marginal likelihood
$\pi(\bm{y}|\bm{\gamma},g)$ facilitates the development of Metropolis-within-Gibbs algorithms that simulate
$\bm{\gamma}$ from $\pi(\bm{\gamma}|g,\bm{y})$ and $g$ from $\pi(g|\bm{\gamma},\bm{y})$ when $p$ is greater than 30.
The MCMC sample of the $\bm{\gamma}$'s is then used to estimate the posterior distribution of a quantity of interest $\Delta$ by Bayesian model averaging
\[\pi(\Delta|\bm{y})=\sum_{\bm{\gamma}}\pi(\Delta|\bm{\gamma},\bm{y})\;\pi(\bm{\gamma}|\bm{y})\]
through the ergodic averages
\[\hat{\pi}(\Delta|\bm{y})=\frac{1}{T}\sum_{i=1}^T \pi(\Delta|\bm{\gamma}^{(i)},\bm{y}),\]
where $T$ is the MCMC sample size and $\bm{\gamma}^{(i)}$ is the $i$th value drawn by the sampler.

The two most commonly implemented algorithms for sampling $\bm{\gamma}$ are the MC$^3$ and Gibbs algorithms. These algorithms select
a coordinate of $\bm{\gamma}$ at random using uniform selection probabilities $\bm{d}=(1/p,\ldots,1/p)$ and then propose to update that
coordinate. The MC$^3$ algorithm is a Metropolis-Hastings algorithm and  was first proposed by \cite{Madigan.York:95}. \cite{Raftery.Madigan:97} used the
MC$^3$ algorithm in BMA for linear regression and \cite{Feldkircher.Zeugner:09} provide the R package \texttt{BMS} to perform BMA in linear regression using the MC$^3$sampler. This algorithm proceeds as follows:
\begin{algorithm}[MC$^3$]
Let $\bm{\gamma}$ be the current state of the chain at time $t$.
\begin{enumerate}
\item Choose coordinate $i$ of $\bm{\gamma}$ using the uniform selection probabilities $\bm{d}$ and propose the new model
$\bm{\gamma}'=(\gamma_1,\ldots,1-\gamma_i,\ldots,\gamma_p)$.
\item Jump to the model $\bm{\gamma}'$ with probability
\begin{equation*}
\alpha(\bm{\gamma},\bm{\gamma}')=\min \left\{
1,\frac{\pi(\bm{\gamma}'|\bm{y},g)}{\pi(\bm{\gamma}|\bm{y},g)}  \right\}.
\end{equation*}
\end{enumerate}
\end{algorithm}
Alternatively, \cite{George.McCullogh:93} used a Gibbs algorithm to sample from the posterior model distribution. This algorithm has the following form:
\begin{algorithm}[Gibbs]
Let $\bm{\gamma}$ be the current state of the chain at time $t$.
\begin{enumerate}
\item Choose coordinate $i$ of $\bm{\gamma}$ using the uniform selection probabilities $\bm{d}$.
\item Generate $\delta\sim\mbox{Bernoulli}\left(\displaystyle\frac{p_i}{1+p_i}\right)$, where
\[p_i=\frac{\pi(\bm{y}|\gamma_i=1,\bm{\gamma}_{-i},g)\;\pi(\gamma_i=1,\bm{\gamma}_{-i})}{\pi(\bm{y}|\gamma_i=0,\bm{\gamma}_{-i},g)\;\pi(\gamma_i=0,\bm{\gamma}_{-i})},\] 
$\bm{\gamma}_{-i}=(\gamma_1,\ldots,\gamma_{i-1},\gamma_{i+1},\ldots,\gamma_p)$ is the vector $\bm{\gamma}$ without the $i$th component and
set $\bm{\gamma}^{(t+1)}=(\gamma_1,\ldots,\gamma_{i-1},\delta,\gamma_{i+1},\ldots,\gamma_p)$.
\end{enumerate}
\end{algorithm}

It is a poor strategy  to use uniform selection probabilities in the large $p$ setting  because  there are many redundant variables which are assigned the same probability as the more important variables. Therefore, the algorithm wastes computational time in proposing poor proposals which results in an inefficient exploration of the model space. Thus, we develop adaptive versions of the MC$^3$ and Gibbs algorithms that update the variable selection probabilities $\bm{d}$ during the simulation in an attempt to automatically decrease the selection probabilities of the many redundant variables. 

\section{Adaptive MCMC Algorithms for BMA in the Linear Regression Model}
Some information about the importance of each variable is progressively gathered as the MCMC sampler runs and we can ideally use this to update the
selection probabilities $\bm{d}$. Let $\bm{w}_t=(w_{t1},\ldots,w_{tp})$ be a descriptive measure about the importance of each variable contained in the current MCMC sample of size $t$. Each coordinate of $\bm{w}_t$ is positive and smaller values correspond to variables which are more likely to be redundant. We define the variable selection probabilities $\bm{d}_t=(d_{t1},\dots,d_{tp})$ at iteration $t$ as follows
\begin{equation} d_{ti}\propto(1-\varepsilon)w_{ti}+\varepsilon, \qquad i=1,\ldots,p,\label{eqn:adaptive_selection_prob}\end{equation}
for $0<\varepsilon<1$. The selection probabilities are non-zero so each $\gamma_i$ could be updated at each iteration, but redundant variables have smaller selection probabilities. The form of selection probabilities in (\ref{eqn:adaptive_selection_prob}) is proportional to a mixture
of a discrete distribution  depending on the current MCMC sample and a uniform distribution. The Adaptive Metropolis algorithm of \cite{Roberts.Rosenthal:09} uses a quite similar mixture distribution  with two multivariate normal component distributions. The covariance matrix of one of the components depends on the empirical variance  of the current MCMC sample while the other has fixed parameters. This mixture proposal distribution is used to ensure the ergodicity of the Adaptive Metropolis algorithm.

Two possible and simple choices of $\bm{w}_t$ are the sample variances $\bm{s}_t^2=(s_{t1}^2,\ldots,s_{tp}^2)$
and the inclusion frequencies $\bm{m}_t=(m_{t1},\ldots,m_{tp})$ of the  coordinates of $\bm{\gamma}$. The coordinates of the sample of $\bm{\gamma}$'s corresponding to redundant variables will be zero in most of the cases and therefore their sample variances  and their inclusion frequencies  will be low leading to small selection probabilities for those redundant variables. On the other hand, important variables are usually in or out of the visited models in most variable selection problems with large $p$ and therefore their coordinates have larger variance and much larger inclusion frequencies leading to higher selection probabilities for those variables. The adaptive MC$^3$ algorithm (denoted by ADMC$^3$) with $\bm{w}_t$ the
sample variances $\bm{s}_t^2$ proceeds as follows:\\
\begin{algorithm}[ADMC$^3(s^2)$]
Let $\bm{\gamma}$ be the current state of the chain at time $t$.
\begin{enumerate}
\item Compute the sample variances $\bm{s}_t^2$ of the coordinates of $\bm{\gamma}$.
\item Choose coordinate $i$ of $\bm{\gamma}$ using selection probabilities $\bm{d}_t$ in (\ref{eqn:adaptive_selection_prob}) with $\bm{w}_t=\bm{s}_t^2$ and propose the new model
$\bm{\gamma}'=(\gamma_1,\ldots,1-\gamma_i,\ldots,\gamma_p)$.
\item Jump to the model $\bm{\gamma}'$ with probability
\begin{equation*}
\alpha(\bm{\gamma},\bm{\gamma}')=\min \left\{
1,\frac{\pi(\bm{\gamma}'|\bm{y},g)}{\pi(\bm{\gamma}|\bm{y},g)}  \right\}.
\label{eqn:MC^3_acceptance_prob}
\end{equation*}
\end{enumerate}
\end{algorithm}
This adaptive algorithm retains the simple implementation of the original MC$^3$ algorithm because a single and easily computed step is added in the MC$^3$ algorithm. The Gibbs algorithm can also be made adaptive in the same way. The pseudocode representation of the adaptive Gibbs algorithm (denoted by ADGibbs) has the following form

\begin{algorithm}[ADGibbs$(s^2)$]
Let $\bm{\gamma}$ be the current state of the chain at time $t$.
\begin{enumerate}
\item Compute the sample variances $\bm{s}_t^2$ of the coordinates of $\bm{\gamma}$.
\item Choose coordinate $i$ of $\bm{\gamma}$ using selection probabilities $\bm{d}_t$  in (\ref{eqn:adaptive_selection_prob}) with $\bm{w}_t=\bm{s}_t^2$.
\item Generate $\delta\sim\mbox{Bernoulli}\left(\displaystyle\frac{p_i}{1+p_i}\right)$, where
\[p_i=\frac{\pi(\bm{y}|\gamma_i=1,\bm{\gamma}_{-i},g)\;\pi(\gamma_i=1,\bm{\gamma}_{-i})}{\pi(\bm{y}|\gamma_i=0,\bm{\gamma}_{-i},g)\;\pi(\gamma_i=0,\bm{\gamma}_{-i})},\] 
$\bm{\gamma}_{-i}=(\gamma_1,\ldots,\gamma_{i-1},\gamma_{i+1},\ldots,\gamma_p)$ is the vector $\bm{\gamma}$ without the $i$th component and set
$\bm{\gamma}^{(t+1)}=(\gamma_1,\ldots,\gamma_{i-1},\delta,\gamma_{i+1},\ldots,\gamma_p)$.

\end{enumerate}
\end{algorithm}
The adaptive Gibbs algorithm also retains the computational simplicity of the original Gibbs algorithm for sampling the model space of BMA in linear regression problems. If we replace  the sample variances $\bm{s}_t^2$ with the sample mean $\bm{m}_t$ of $\bm{\gamma}$ in Steps 1 and 2 of both algorithms then we get the ADMC$^3(m)$ and ADGibbs$(m)$ algorithms. Alternative descriptive sample  measures, which involve extra computational cost,  include the Rao-Blackwellized estimates of variable inclusion probabilities discussed in \cite{Guan.Stephens:11}.

In practice, we split the iterations in $B$ blocks of $l$ iterations and update the adaptive Step 1 at the end of each block. We also set $\varepsilon=1/p$ to perform a more considerable adaptation in  variable selection problems with large $p$. Alternatively, we could choose to progressively decrease $\varepsilon$ with time because more and more information
accumulates for $\bm{w}_t$. This can be achieved by defining an $\varepsilon_b$ for each block  $b$ as 
$\varepsilon_b={1}/(bp), b=1,\ldots,B.$

\cite{Richardson.Bottolo:10} propose a similar adaptive scanning strategy in multivariate regression analysis that aims to increase the probability of updating the more interesting responses among a large number of them. The more interesting  responses are those that are more likely to be  associated with several predictors. However, to sample $\bm{\gamma}$, they apply an Evolutionary Monte Carlo scheme described in \cite{Bottolo.Richardson:10} with a population of $L$ Markov chains that are simulated in parallel with different temperatures.
\cite{Peltola.Martttinen:12} also propose a quite similar adaptive Metropolis-Hastings algorithm for variable selection that uses estimates of variable inclusion probabilities to continuously update the selection probabilities. However, their adaptive algorithm is finite and this raises an issue on when to stop the adaptive phase whereas our algorithm is performing an infinite adaptation. Moreover,  to bound the selection probabilities away from zero they preselect a minimum value while we have used the mixture distribution (\ref{eqn:adaptive_selection_prob}).
\cite{nottkohn05}  also propose an adaptive Gibbs algorithm for BMA in linear regression that approximates the full conditionals $\pi(\gamma_i=1|\bm{\gamma}_{-i},\bm{y})$ through an easily computed adaptive best linear predictor. However, in contrast to our adaptive methods, they use uniform selection probabilities to choose coordinates $i$ of $\bm{\gamma}$ to perform the model update step.

We also use a simple random walk Metropolis-Hastings step to sample $g$ with a Log-Normal proposal centred over the previous value. The random walk Metropolis-Hastings step could be made adaptive by controlling the scale parameter of the Log-Normal proposal to result in an acceptance rate
equal to 0.44. This automatic tuning of the scale parameter is done in a similar way to the tuning of the scale parameter of the Adaptive Random Walk Metropolis algorithm proposed by  \cite{Atchade.Rosenthal:05}.

\subsection{Ergodicity of the Adaptive MCMC algorithms}

The Gibbs and MC$^3$ samplers  are finite, irreducible and aperiodic for a fixed choice of $\bm{d}$ and therefore
they are uniformly ergodic. Furthermore, the change $|d_{ti}-d_{(t-1)i}|$ in each coordinate of $\bm{d}_t$ converges to 0 as $t$ goes to infinity because the empirical estimates $\bm{w}_t$ are modified by order $\mbox{O}(1/t)$ at iteration $t$. Therefore, Theorem 4.1 of \cite{LatRos10} ensures immediately that the
proposed adaptive (random scan) Gibbs algorithms are ergodic.

Conditions for the ergodicity of any other type of adaptive MCMC algorithm were discussed in \cite{Roberts.Rosenthal:07} who established two sufficient conditions: the simultaneous uniform ergodicity condition and the diminishing adaptation condition. The adaptive MC$^3$ algorithms satisfy the simultaneous uniform
ergodicity condition since the state space $\mathcal{X}=\left\{0,1\right\}^p$ is finite and its proposal density $q_d(\bm{\gamma}'|\bm{\gamma})=d_i$ is continuous in the closed space of selection probabilities.
The diminishing adaptation requires that the amount of adaptation diminishes at each iteration, which is achieved because the transition kernel $P_d(\bm{\gamma}'|\bm{\gamma})$ is continuous with respect to $\bm{d}$ and the selection probabilities $\bm{d}_t$ are modified by order $\mbox{O}(1/t)$ at iteration $t$. More details about the ergodicity of the adaptive MC$^3$ algorithms are presented in the Appendix.

\section{Illustrations}
The performance of the MCMC algorithms is evaluated using simulated datasets and two real datasets from cross-country growth regressions.
The simulated datasets use the $n\times p$ matrix $\bm{Z}$ implemented in example 4.2 of \cite{George.McCullogh:93}. The columns of $\bm{Z}$
are generated in the following way
\[ \bm{z}_i=\bm{z}_i^{*}+\bm{e}\]
where $\bm{z}_i^{*}$ and $\bm{e}$ are vectors of $n$ independent standard normal elements and therefore the components of $\bm{Z}$  have pairwise correlation of 0.5.
After demeaning to obtain the design matrix $\bm{X}$ we generate $n$ observations  from Model 3 discussed in \cite{Ley.Steel:08} as follows
\[\bm{y}=\bm{1}+\sum_{i=1}^{7}\bm{x}_i+\tau \bm{v}\]
where $\bm{v}$ is a vector of $n$ independent standard normal elements and we set $\tau=2$. Values of $p$ used are 40 and 80 while we adopt $n=50$ to consider the cases $p<n$ and $p>n$. Finally, we have simulated five different datasets for each value of $p$.

The first real dataset was used in \cite{Fernandez.et.al:01} (FLS) and contains $p=41$ determinants of economic growth for $n=72$ countries
whereas the second dataset was introduced by \cite{Sala-i-Martin:04} (SDM) and contains $p=67$ determinants of economic growth for $n=88$ countries.

All the  MCMC samplers  were run for 2,000,000 iterations with a burn-in period of 100,000 iterations
and thinned every 10th iteration resulting in an MCMC sample size $T$ of 190,000. We choose mean prior model size $\kappa=7$ and use both the Hyper$-g/n$ prior and the Benchmark $g-$prior ($g-$BRIC). In the adaptive samplers, the variable selection probabilities $\bm{d}_t$ are updated  every 1,000 thinned MCMC samples and adaptation starts after the first 10,000 thinned samples (i.e.~at block index $b=10$) with total number of blocks $B=190$.

The efficiency of an MCMC sampler can be measured using the Effective Sample Size (ESS) which is $T/(1+2\sum_{j=1}^{\infty}
\rho_j)$ for an MCMC run of length $T$ with lag $j$ autocorrelation $\rho_j$ \citep[{\it e.g.},][]{Liu:01}. The interpretation is that
the MCMC sampler leads to the same accuracy of estimates as a Monte Carlo sampler (where all the draws are independent) run for ESS
iterations. A quite important posterior measure in those problems is the posterior variable inclusion probability (PIP) and therefore the MCMC output monitored in this paper  consists of those components $\gamma_i$ of $\bm{\gamma}$ having PIP greater or equal to 0.1 (a non-negligible inclusion probability for those problems). The same variables were found to have posterior inclusion probability  greater or equal to 0.1  in all algorithms  for each prior setting on regression coefficients and dataset. An estimate of the integrated autocorrelation time $\tau_i= 1+2\sum_{j=1}^{\infty} \rho_j$ for each $\gamma_i$ with PIP$\geq 0.1$ was
computed using the Lag Window Estimator \citep{Geyer:92} with a Parzen window kernel. We calculate the median $M$ of $\tau_i$'s  for each
algorithm and estimate the Effective Sample Size by $\mbox{ESS}=T/M$.
The algorithms have different running times and so we also define the efficiency ratio for a sampler to be
\[
\mbox{ER(Sampler)}=\frac{\mbox{ESS(Sampler)}}{\mbox{CPU(Sampler)}},
\]
which standardizes the ESS  by CPU run time and so
penalizes computationally demanding algorithms. We are also interested
in the performance of each adaptive algorithm relative to the non-adaptive
algorithm and the relative efficiency of the adaptive over
the non-adaptive algorithm is defined by
\begin{align*}
\mbox{RE}=&\frac{\mbox{ER(Adaptive)}}{\mbox{ER(Non-Adaptive)}}.
\end{align*}
Adaptive versions of the MC$^3$ and Gibbs algorithms are denoted as in Section 3.

\begin{table}[h!]
\caption{\small The effective sample size ESS, the CPU time in
seconds, the efficiency ratio ER of the non-adaptive and adaptive algorithms  with relative
efficiencies RE of the adaptive algorithm over the non-adaptive
algorithm and the acceptance rate $\tilde{A}$ for the simulated datasets and $g-$BRIC prior }
\centerline{\small
\renewcommand{\baselinestretch}{1.2}
\begin{tabularx}{\linewidth}{XXXXXX}\\
\hline   Method                    &    ESS           &  CPU          & ER            &    RE             & $\tilde{A}$ \\ \hline
                                   &  &\multicolumn{2}{c}{$p=40$}\\
         MC$^3$                    & 11756 (1000)            & 2963 (138)          &  4.06 (0.39)         &                & 7\%    \\
         ADMC$^3(s^2)$             & 31370 (2587)            & 2894 (140)          &  {\bf 11.53} (1.43)        &  2.40 (0.11)   & 21\%   \\
         ADMC$^3(m)$               & 19872 (1275)            & 2861 (142)          &  7.18 (0.57)         &  1.84 (0.06)   & 18\%   \\
         Gibbs                     & 8830 (795)              & 3362 (143)          &  2.70 (0.28)         &                &\\
         ADGibbs$(s^2)$            & 22415 (1720)   				 & 3349 (146)          &  7.04 (0.80)         &  2.26 (0.07)   &\\
         ADGibbs$(m)$              & 14771 (1089)            & 3333 (142)          &  4.54 (0.39)         &  1.80 (0.09)   &\\
                                   &  &\multicolumn{2}{c}{$p=80$}\\
         MC$^3$                    & 4972 (289)             & 2737 (81)            &  1.90 (0.16)         &                & 3\%   \\
         ADMC$^3(s^2)$             & 12804 (706)            & 2610 (73)            &  {\bf 5.11} (0.42)         &  2.76 (0.11)   & 20\%  \\
         ADMC$^3(m)$               & 8775 (841)             & 2553 (70)            &  3.59 (0.42)         &  1.86 (0.13)   & 16\%  \\
         Gibbs                     & 3556 (197)             & 3074 (57)            &  1.18 (0.08)         &                &\\
         ADGibbs$(s^2)$            & 10887 (715)            & 3101 (73)            &  3.64 (0.31)         &  3.07 (0.13)   &\\
         ADGibbs$(m)$              & 7870 (559)             & 3043 (65)            &  2.68 (0.24)         &  2.22 (0.07)   &\\ \hline
\end{tabularx}}
\label{table:simulData_efficiency_adaptive_gBRIC}
\end{table}
\renewcommand{\baselinestretch}{1.5}

\begin{table}[h!]
\caption{\small The effective sample size ESS, the CPU time in
seconds, the efficiency ratio ER of the non-adaptive and adaptive algorithms  with relative
efficiencies RE of the adaptive algorithm over the non-adaptive
algorithm and the acceptance rate $\tilde{A}$ for the FLS and SDM datasets and $g-$BRIC prior }
\centerline{\small
\renewcommand{\baselinestretch}{1.2}
\begin{tabularx}{\linewidth}{XXXXXX}\\
\hline   Method                    &    ESS           &  CPU          & ER            &    RE     & $\tilde{A}$ \\ \hline
                                   & &\multicolumn{2}{c}{FLS data}\\
         MC$^3$                    & 7759             & 4181          &  1.86         &           & 8\%\\
         ADMC$^3(s^2)$             & 14218            & 4108          & {\bf 3.46}         &  1.87     & 18\%\\
         ADMC$^3(m)$               & 13755            & 4054          &  3.39         &  1.83     & 15\%\\
         Gibbs                     & 5778             & 4616          &  1.25         &           &\\
         ADGibbs$(s^2)$            & 10983            & 4621          &  2.38         &  1.90     &\\
         ADGibbs$(m)$              & 10087            & 4611          &  2.29         &  1.75     &\\                           
                                   & &\multicolumn{2}{c}{SDM data}\\
         MC$^3$                    & 1998             & 2560          &  0.78         &           & 2\% \\
         ADMC$^3(s^2)$             & 9056             & 2464          &  {\bf 3.68}         &  4.71     & 7\% \\
         ADMC$^3(m)$               & 5735             & 2379          &  2.42         &  3.10     & 5\%  \\
         Gibbs                     & 1829             & 3037          &  0.60         &           &\\
         ADGibbs$(s^2)$            & 7914             & 3021          &  2.62         &  4.35     &\\
         ADGibbs$(m)$              & 5364             & 2893          &  1.85         &  3.08     &\\ \hline
\end{tabularx}}
\label{table:efficiency_adaptive_gBRIC}
\end{table}
\renewcommand{\baselinestretch}{1.5}

Table \ref{table:simulData_efficiency_adaptive_gBRIC} and Table \ref{table:efficiency_adaptive_gBRIC} present results of the adaptive and non-adaptive samplers for the simulated and real datasets respectively and the $g-$BRIC prior setting. Standard errors for the estimates over the five different simulated datasets for each value of $p$ are also provided in Table \ref{table:simulData_efficiency_adaptive_gBRIC}. The ADMC$^3(s^2)$ and ADGibbs$(s^2)$ samplers tend to have the highest ESS, followed by ADMC$^3(m)$ and ADGibbs$(m)$ and finally the MC$^3$  and Gibbs samplers. The adaptive MC$^3$ samplers are more efficient than the adaptive Gibbs one. Furthermore, the adaptive samplers that use the vector of sample variances $\bm{s}^2$ to update the variable selection probabilities consistently outperform their adaptive counterparts that use the inclusion frequencies $\bm{m}$. If we take computing time into account, the order of the samplers in terms of their efficiency remains almost the same. Best performance in terms of efficiency ratio is indicated by bold numbers.

The RE of the adaptive algorithms over the non-adaptive algorithms are always greater than 1 indicating that  adaptive methods are superior.
The most benefit from adaptation appears in the simulated dataset with $p=80$ and the SDM dataset which are those datasets with the larger number of variables. The  ADMC$^3(s^2)$ and ADGibbs$(s^2)$ are almost three times more efficient from their non-adaptive counterparts for the simulated dataset with $p=80$ while they are more than four times more efficient for the SDM data. Therefore, adaptation tends to provide more efficiency in datasets with large number of variables (more than 40 variables). As many of these are redundant, this is what we would expect. Finally, the adaptive MC$^3$ algorithms always have much more reasonable (higher) between-model acceptance rates and this should lead to a more efficient exploration of the model space.

\begin{table}[h!]
\caption{\small The effective sample size ESS, the CPU time in
seconds, the efficiency ratio ER of the non-adaptive and adaptive algorithms  with relative
efficiencies RE of the adaptive algorithm over the non-adaptive
algorithm and the acceptance rate $\tilde{A}$ for the simulated datasets and Hyper$-g/n$ prior }
\centerline{\small
\renewcommand{\baselinestretch}{1.2}
\begin{tabularx}{\linewidth}{XXXXXX}\\
\hline   Method                    &    ESS           &  CPU          & ER            &    RE                             & $\tilde{A}$ \\ \hline
                                   & & \multicolumn{2}{c}{$p=40$}\\
        MC$^3$                     & 15661 (1189)             & 5565 (194)          &  2.91 (0.27)         &              & 14\%  \\
         ADMC$^3(s^2)$             & 24490 (1487)             & 5542 (190)          &  {\bf 4.39} (0.37)         &  1.51 (0.02) & 22\% \\
         ADMC$^3(m)$               & 20349 (911)              & 5558 (213)          &  3.79 (0.25)         &  1.38 (0.07) & 21\%\\
         Gibbs                     & 11543 (858)              & 6112 (223)          &  1.97 (0.19)         &              &\\
         ADGibbs$(s^2)$            & 18778 (1204)             & 6038 (185)          &  3.20 (0.26)         &  1.66 (0.02) &\\
         ADGibbs$(m)$              & 15972 (725)              & 6035 (158)          &  2.69 (0.15)         &  1.46 (0.06) &\\
                                   & &\multicolumn{2}{c}{$p=80$}\\
         MC$^3$                    & 6033 (290)              & 5372 (74)           &  1.14 (0.07)         &               & 6\%   \\
         ADMC$^3(s^2)$             & 14893 (459)             & 5262 (84)           &  {\bf 2.85} (0.10)         &  2.60 (0.11)  & 16\%  \\
         ADMC$^3(m)$               & 11585 (320)             & 5197 (95)           &  2.25 (0.08)         &  2.07 (0.10)  & 16\%  \\
         Gibbs                     & 4714 (206)              & 5730 (78)           &  0.83 (0.04)         &               &\\
         ADGibbs$(s^2)$            & 11097 (237)             & 5790 (103)          &  1.94 (0.07)         &  2.42 (0.12)  &\\
         ADGibbs$(m)$              & 10319 (615)             & 5834 (105)          &  1.80 (0.13)         &  2.19 (0.12)  &\\ \hline
\end{tabularx}}
\label{table:simulData_efficiency_adaptive_Hyper_g_prior}
\end{table}
\renewcommand{\baselinestretch}{1.5}

\begin{table}[h!]
\caption{\small The effective sample size ESS, the CPU time in
seconds, the efficiency ratio ER of the non-adaptive and adaptive algorithms  with relative
efficiencies RE of the adaptive algorithm over the non-adaptive
algorithm and the acceptance rate $\tilde{A}$ for the FLS and SDM datasets and Hyper$-g/n$ prior }
\centerline{\small
\renewcommand{\baselinestretch}{1.2}
\begin{tabularx}{\linewidth}{XXXXXX}\\
\hline   Method                    &    ESS           &  CPU          & ER              &    RE     & $\tilde{A}$ \\ \hline
                                   & & \multicolumn{2}{c}{FLS data}\\
        MC$^3$                     & 13001             & 14337          &  0.91         &           & 25\%\\
         ADMC$^3(s^2)$             & 14003             & 14254          &  0.98         &  1.08     & 30\%\\
         ADMC$^3(m)$               & 17921             & 14357          &  {\bf 1.25}      &  1.38     & 26\%\\
         Gibbs                     & 8997              & 15203          &  0.59         &           &\\
         ADGibbs$(s^2)$            & 10106             & 15366          &  0.66         &  1.11     &\\
         ADGibbs$(m)$              & 11472             & 15155          &  0.76         &  1.28     &\\                           
                                   & & \multicolumn{2}{c}{SDM data}\\
         MC$^3$                    & 5480             & 10061          &  0.54         &            & 13\% \\
         ADMC$^3(s^2)$             & 9487             & 9993           &  {\bf 0.95}         &  1.74      & 17\%\\
         ADMC$^3(m)$               & 9219             & 9860           &  0.94         &  1.72      & 19\%\\
         Gibbs                     & 4617             & 10624          &  0.43         &            & \\
         ADGibbs$(s^2)$            & 7334             & 10631          &  0.69         &  1.59      & \\
         ADGibbs$(m)$              & 7276             & 10708          &  0.68         &  1.56      & \\ \hline
\end{tabularx}}
\label{table:efficiency_adaptive_Hyper_g_prior}
\end{table}
\renewcommand{\baselinestretch}{1.5}

Table \ref{table:simulData_efficiency_adaptive_Hyper_g_prior} and Table \ref{table:efficiency_adaptive_Hyper_g_prior} display the results of the adaptive and non-adaptive samplers for the simulated and real datasets respectively and the Hyper$-g/n$ prior setting. The adaptive MC$^3$ samplers tend to have the highest ESS, followed by the adaptive Gibbs algorithms and finally the MC$^3$ and Gibbs samplers. The adaptive algorithms using the sample variances $\bm{s}^2$ to update the selection probabilities again tend to perform better than those using the sample inclusion frequencies $\bm{m}$ (except for the FLS data). 
Finally, the rank of the samplers remain unchanged if we take computing time into consideration.

The adaptive algorithms are more efficient than the non-adaptive ones because the RE of the adaptive over the non-adaptive algorithms are greater than one. Adaptation again tends to be more effective in datasets with  larger number of variables and results in higher between-model acceptance rate than the non-adaptive algorithms.

The sample variances $\bm{s}^2$ are generally  better descriptive measures for updating the variable selection probabilities than the inclusion probabilities that were recently used in the adaptive Metropolis-Hastings algorithm of \cite{Peltola.Martttinen:12}. Overall, the adaptive MC$^3$ algorithm that use the sample variances $\bm{s}^2$ to update the selection probabilities (the ADMC$^3(s^2)$ sampler) seems to be the most efficient algorithm and it is the one recommended in this study, particularly with large $p$ (more than 40 variables).

Table \ref{table:proportion_regressors_non_negligible_PIP} presents the mean number $k$ of variables with PIP$\geq 0.1$ (a non-negligible PIP for those problems) and the proportion $k/p$ for each prior setting and dataset. The number $k$ is computed as the average over all considered algorithms. In line with expectation, there seems to be an association between the proportion of variables with non-negligible PIP and the efficiency gain of adaptation. For example, in the case of the SDM dataset with $g-$BRIC prior, the proportion of variables with PIP$\geq 0.1$ is very low (it is equal to 0.06) and the adaptive algorithms are over four times more efficient than non-adaptive algorithms. On the other hand, in the case of the FLS dataset with Hyper$-g/n$ prior, $k/p$ is quite large (it is equal to 0.40) and the adaptive algorithms are only marginally better than non-adaptive algorithms. Therefore, the smaller the proportion of variables with PIP$\geq 0.1$, the higher the efficiency gain of adaptation. Intuitively, adaptation can make a lot of difference where there are many unimportant variables as for those variables the proposal probabilities can be made quite small. This will increase the acceptance rate of proposed models and the efficiency of the algorithms. 

In the large $p$ setting, the $g-$BRIC prior results in a (sometimes much) smaller number of variables with PIP$\geq 0.1$ than the Hyper$-g/n$ prior because it induces a higher model size penalty. Large values of $g$ increase the model size penalty \citep{Ley.Steel:08} and in our simulated and real datasets the value of $g=p^2$ (implied by the $g-$BRIC prior) is much larger than the posterior median of $g$ under the Hyper$-g/n$ prior. Therefore, the
$g-$BRIC prior induces a  higher model size penalty and leads to a smaller proportion of variables with PIP$\geq 0.1$. This explains why the $g-$BRIC prior tends to gain more benefit from adaptation.

\begin{table}[h!]
\caption{\small Number and proportion of regressors with PIP$\geq 0.1$ for each prior setting and dataset}
\begin{center}\begin{tabular}{ll@{\hskip 0.5cm}c@{\hskip 0.5cm}@{\hskip 0.5cm}c@{\hskip 0.5cm}}
\toprule
Prior setting        &      Dataset               &     $k$        & $k/p$   \\ \toprule
                     &     Simulated $(p=40)$     &      8         &  0.20   \\
$g-$ BRIC            &     Simulated $(p=80)$     &      8         &  0.10   \\
                     &     FLS                    &      10        &  0.24   \\
                     &     SDM                    &      4         &  0.06   \\  \toprule                                
                     &     Simulated $(p=40)$     &      12        &  0.30   \\
Hyper$-g/n$          &     Simulated $(p=80)$     &      9         &  0.11   \\
                     &     FLS                    &      27        &  0.40   \\
                     &     SDM                    &      23        &  0.34  \\   \bottomrule
\label{table:proportion_regressors_non_negligible_PIP}
\end{tabular}
\end{center}
\end{table}

\begin{figure}[h!]
\begin{center}
\centerline{$g-$BRIC}
\includegraphics[scale=0.45]{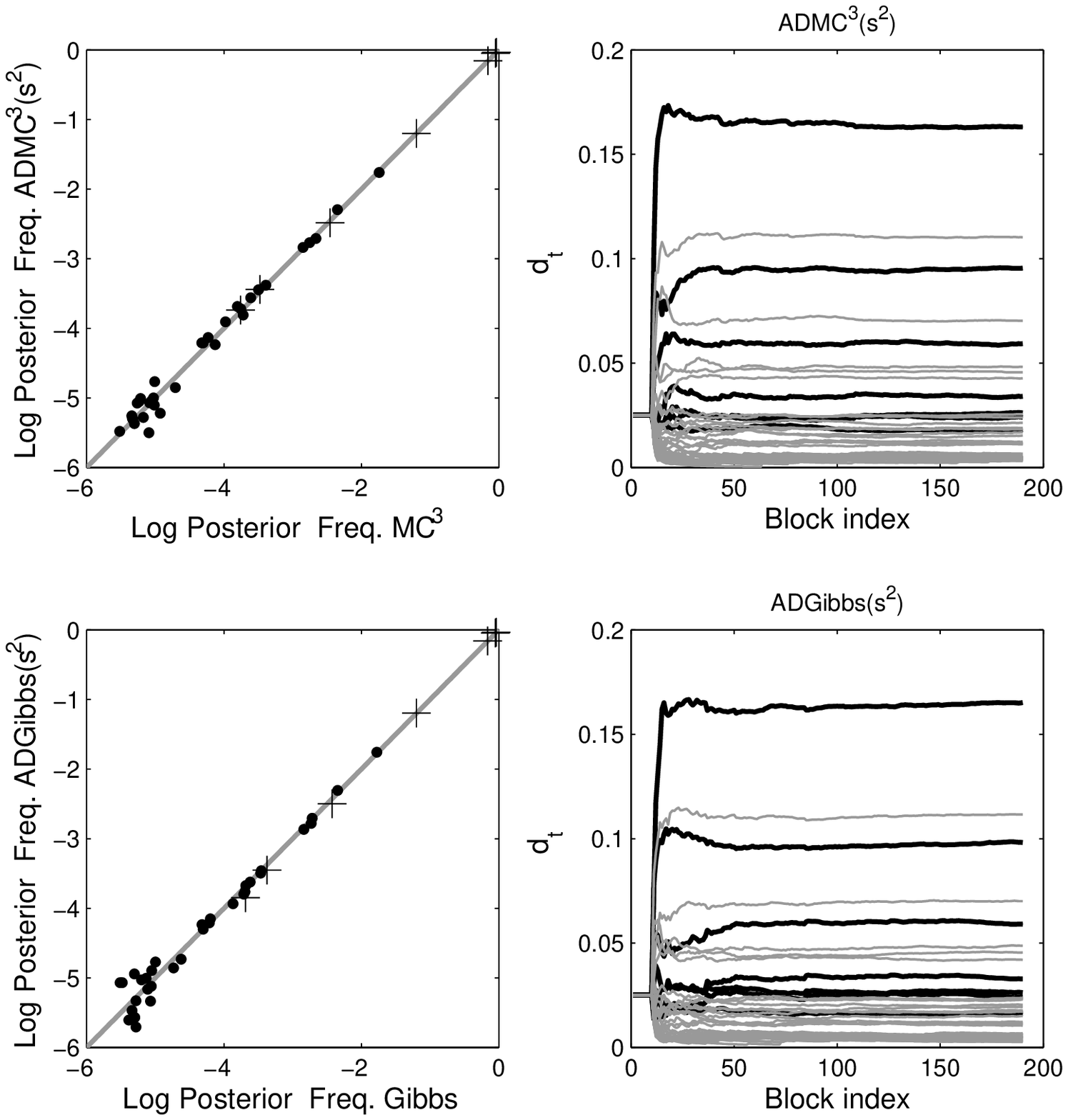}
\centerline{Hyper$-g/n$ prior}
\includegraphics[scale=0.45]{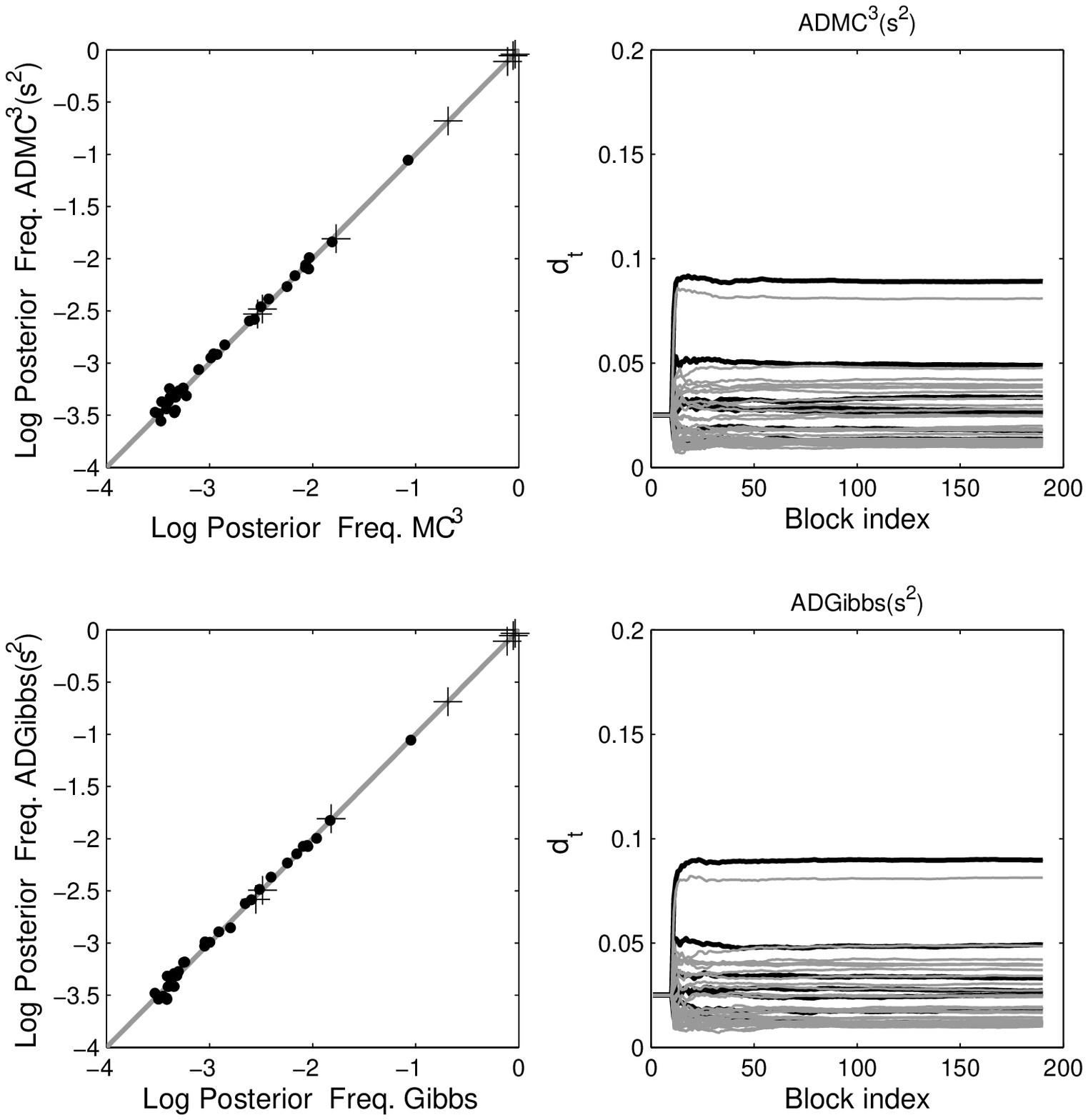}
\end{center}
\caption{\small Scatter-plot of log estimated posterior inclusion probabilities of the adaptive and non-adaptive algorithms (+ denotes variables included in the true model) and trace plot of selection probabilities  for ADMC$^3(s^2)$ and ADGibbs$(s^2)$  algorithms (black: true variables, light grey: redundant variables) for each prior setting and $p=40$ }
\label{fig:Adaptive_vs_nonadaptive_post_freq_p_40}
\end{figure}

\begin{figure}[h!]
\begin{center}
\centerline{$g-$BRIC}
\includegraphics[scale=0.45]{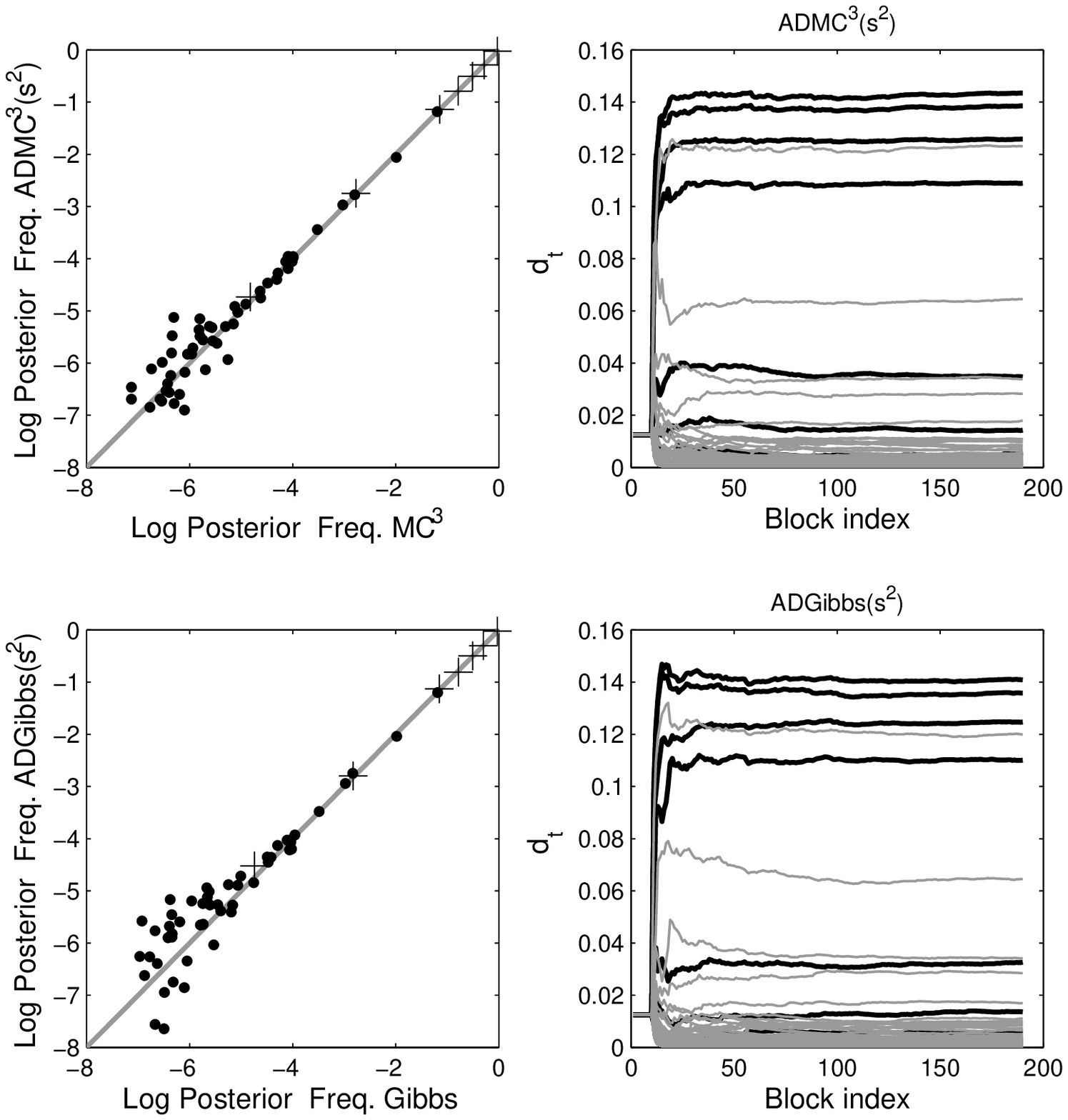}
\centerline{Hyper$-g/n$ prior}
\includegraphics[scale=0.45]{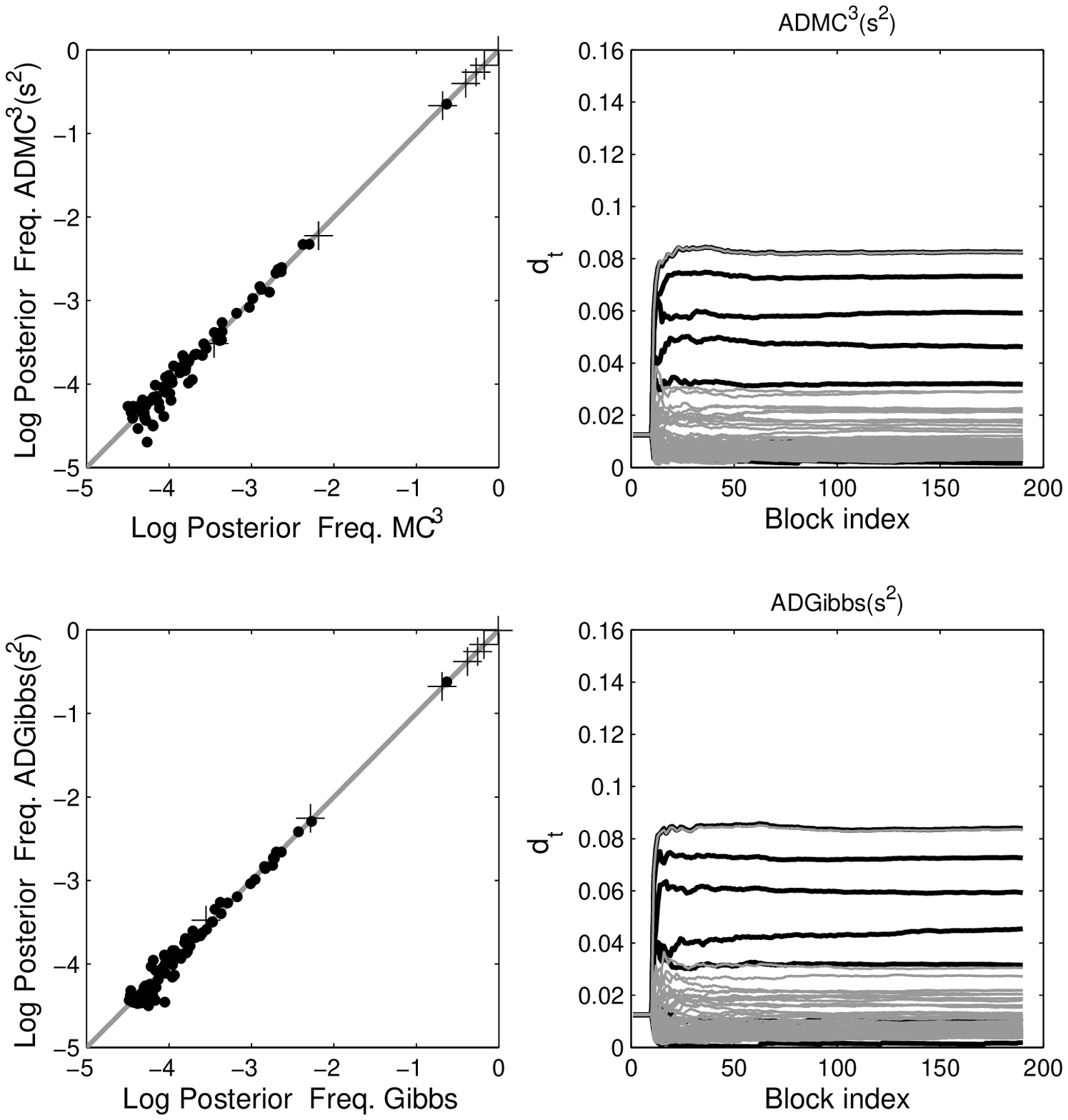}
\end{center}
\caption{\small Scatter-plot of log estimated posterior inclusion probabilities of the adaptive and non-adaptive algorithms (+ denotes variables included in the true model) and trace plot of selection probabilities  for ADMC$^3(s^2)$ and ADGibbs$(s^2)$  algorithms (black: true variables, light grey: redundant variables) for each prior setting and $p=80$  }
\label{fig:Adaptive_vs_nonadaptive_post_freq_p_80}
\end{figure}

Figure \ref{fig:Adaptive_vs_nonadaptive_post_freq_p_40} and Figure \ref{fig:Adaptive_vs_nonadaptive_post_freq_p_80} display the scatter-plot of log estimated PIP of the adaptive and non-adaptive algorithms (left column) and the trace plot of selection probabilities for ADMC$^3(s^2)$ and ADGibbs$(s^2)$ algorithms (right column)  for each prior setting and $p=40$ and 80 respectively. The posterior variable inclusion probabilities are very similar for the adaptive and non-adaptive algorithms and they are almost identical for those variables with PIP$\geq 0.01$ ($\log(\mbox{PIP})\geq -4.6$) and the true variables (shown with $+$). However, the adaptive algorithms gain efficiency by adjusting the uniform  selection probabilities and assigning a much lower variable selection probability to redundant variables, as illustrated in the right panels of Figures \ref{fig:Adaptive_vs_nonadaptive_post_freq_p_40} and \ref{fig:Adaptive_vs_nonadaptive_post_freq_p_80}.

The variable selection probabilities are adapted more in the $g-$BRIC prior setting, which induces a higher model size penalty and concentrates the posterior model distribution on parsimonious models. Thus, many redundant variables have  lower selection probabilities and a few variables have higher selection probabilities than under the Hyper$-g/n$ prior. Therefore, redundant variables are proposed less often with the $g-$BRIC prior and adaptation tends to be more effective in this prior setting. 

\setlength{\tabcolsep}{0.6em}
\begin{table}[h!]
\caption{\small FLS data-Marginal posterior inclusion probabilities of some economic growth determinants}
\begin{footnotesize}
\begin{tabular}{lcccccc}
\hline
                                     &\multicolumn{3}{c}{$g-$BRIC} &\multicolumn{3}{c}{Hyper$-g/n$} \\ 
Regressors                           &MC$^3$&ADMC$^3(s^2)$&ADMC$^3(m)$&MC$^3$&ADMC$^3(s^2)$&ADMC$^3(m)$\\ \hline
log GDP in 1960                      &0.70&0.64&0.62&0.99&0.99&0.99\\
Fraction Confucian                   &0.96&0.96&0.96&0.98&0.98&0.98\\
Life expectancy                      &0.45&0.39&0.38&0.84&0.84&0.84\\
Equipment investement                &0.98&0.99&0.99&0.90&0.90&0.90\\
Sub-Saharan dummy                     &0.50&0.47&0.46&0.74&0.74&0.75\\
Fraction Muslim                      &0.19&0.15&0.15&0.53&0.54&0.53\\
Number of years open economy         &0.56&0.52&0.52&0.48&0.48&0.47\\
Degree of capitalism                 &0.09&0.08&0.08&0.50&0.50&0.50\\
Fraction Protestant                 &0.39&0.38&0.39&0.61&0.61&0.62\\
Non-Equipment investment             &0.09&0.08&0.08&0.49&0.50&0.50\\
Fraction GDP mining                  &0.06&0.05&0.05&0.51&0.51&0.51\\\hline
\label{table:Posterior_Inclusion_Probabilities_FLS}
\end{tabular}
\end{footnotesize}
\end{table}

\begin{table}[h!]
\caption{\small SDM data-Marginal posterior inclusion probabilities of some economic growth determinants}
\begin{footnotesize}
\begin{tabular}{lcccccc}
\hline
                                     &\multicolumn{3}{c}{$g-$BRIC} &\multicolumn{3}{c}{Hyper$-g/n$} \\ \cline{2-7}
Regressors                           &MC$^3$&ADMC$^3(s^2)$&ADMC$^3(m)$&MC$^3$&ADMC$^3(s^2)$&ADMC$^3(m)$\\ \hline
East Asian dummy                                    &0.99&0.99&0.99&0.75&0.77&0.76\\
log GDP in 1960                                     &0.01&0.02&0.01&0.65&0.65&0.66\\
Investment price                                    &0.04&0.04&0.04&0.72&0.72&0.74\\
Malaria prevalence in 1960                          &0.83&0.83&0.82&0.27&0.27&0.26\\
Primary schooling in 1960                            &0.17&0.16&0.17&0.74&0.74&0.74\\ \hline
\label{table:Posterior_Inclusion_Probabilities_SDM}
\end{tabular}
\end{footnotesize}
\end{table}

\begin{figure}[h!]
\centerline{
\includegraphics[scale=0.6]{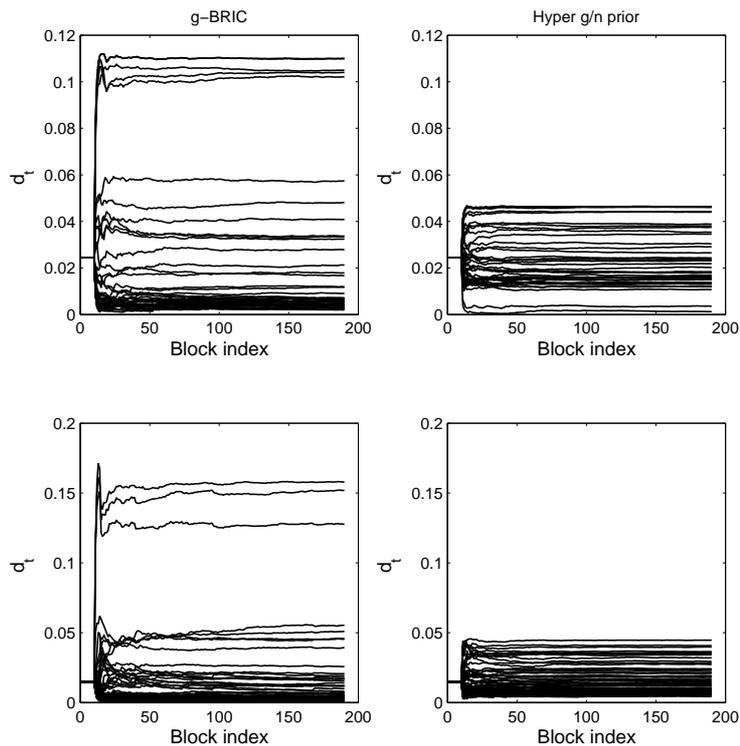}}
\caption{\small Trace plot of selection probabilities of ADMC$^3(s^2)$ algorithm for each prior setting and the FLS data (upper panel) and SDM data (lower panel) }
\label{fig:Weight_selection_prob_Hyper-g}
\end{figure}

Table \ref{table:Posterior_Inclusion_Probabilities_FLS} and Table \ref{table:Posterior_Inclusion_Probabilities_SDM} present the marginal posterior inclusion probabilities of all regressors that receive an inclusion probability of over 50\%
under any  of the prior settings and algorithms for the FLS and SDM data respectively. The posterior inclusion probabilities of adaptive and non-adaptive algorithms are quite similar for both prior settings and data and they are almost identical in the case of Hyper$-g/n$ prior. This suggests empirically that the adaptive and non-adaptive algorithms converge to the same stationary distribution as should be expected from the theory. Figure \ref{fig:Weight_selection_prob_Hyper-g} displays the trace plot of selection probabilities of ADMC$^3(s^2)$ for each prior setting with the real datasets. It is again clear that the adaptive samplers decrease the selection probabilities of many redundant variables and this provides more efficiency in estimating the PIP of the important variables. The variable selection probabilities are again more markedly different from uniform probabilities in the $g-$BRIC prior setting because this prior induces a larger model size penalty and concentrates the posterior model distribution on parsimonious models.

\section{Conclusions}
Adaptive versions of the MC$^3$ and Gibbs samplers for Bayesian Model Averaging in linear regression models are developed to progressively construct better proposals by adapting the vector of variable selection probabilities. The vector of selection probabilities is proportional to a mixture distribution weighting the sample variances $\bm{s}^2$ or the inclusion frequencies $\bm{m}$ of the variables and the uniform distribution. The adaptive samplers automatically decrease the selection probabilities of many redundant variables and this leads to more efficient samplers, particularly when the proportion of variables with small posterior inclusion probabilities is large.  
As the number of variables available in
applications tends to increase, these adaptive algorithms are useful and easily implemented alternatives or complements to the popular MC$^3$ and Gibbs samplers.  The adaptive MC$^3$ algorithm that uses the sample variances $\bm{s}^2$ to update the selection probabilities is found to be the most efficient algorithm in simulated and real datasets  and it is the one recommended in this study, particularly with a large number of variables (more than 40 variables). 

Extensions of the adaptive ideas presented in this study can also be accommodated in Bayesian Model Averaging for logistic regression model with many more variables than observations ($p\gg n$). Such problems typically arise in genome-wide association studies where gene expression data contains hundreds or even thousands of variables. The MC$^3$ algorithm is very inefficient in those datasets because it spends a large amount of time trying to add a variable before proposing to delete a variable. The more general model proposal discussed in \cite{Lamnisos.Griffin:08a} solves this issue and also combines local moves with more global ones by changing a block of variables simultaneously. We are currently combining the adaptive ideas of the present paper with this model proposal to develop efficient samplers for problems with $p\gg n$.

\bibliographystyle{Chicago}
    \bibliography{bib_NOTES}
		
\section*{Appendix}

The proposal density of the adaptive MC$^3$ algorithm is $q_d(\bm{\gamma}'|\bm{\gamma})=d_i$ and the transition kernel is
\[P_d(\bm{\gamma},\bm{\gamma}')=\alpha(\bm{\gamma}',\bm{\gamma})\; d_i \; \bm{1}_{\{\bm{\gamma}'\neq \bm{\gamma}\}}+\sum_{j=1}^p (1-\alpha(\bm{\gamma}_j',\bm{\gamma}))\;d_j\;\bm{1}_{\{\bm{\gamma}'_j= \bm{\gamma}\}}\]
where $\bm{\gamma}_j'=(\gamma_1,\ldots,1-\gamma_j,\ldots,\gamma_p)$. Both $q_d(\bm{\gamma}'|\bm{\gamma})$ and $P_d(\bm{\gamma},\bm{\gamma}')$ are continuous with respect to $\bm{d}$ in the closed space of selection probabilities. The simultaneous uniform ergodicity condition for the adaptive MC$^3$ algorithms follows from Corollary 3 and Lemma 1 of \cite{Roberts.Rosenthal:07}.

The diminishing adaptation of the adaptive MC$^3$ algorithms results from the continuity of $P_d(\bm{\gamma},\cdot)$ with respect to $\bm{d}$ and the fact that the modification in selection probabilities $\bm{d}_t$ converges to 0 as $t$ goes to infinity since the empirical estimates $\bm{w}_t$ are modified by order $O(1/t)$ at iteration $t$.

\end{document}